\documentclass[prl,twocolumn,showpacs,aps,superscriptaddress,floatfix]{revtex4}

\usepackage{graphicx}
\usepackage{bm}
\usepackage{dcolumn}
\usepackage[usenames]{color}
\def\beq{\begin{equation}}
\def\eeq{\end{equation}}
\def\bea{\begin{eqnarray}}
\def\eea{\end{eqnarray}}

\begin{document}
\title{Minimal Uncertainty in Momentum: The Effects of IR
Gravity on Quantum Mechanics}

\author{B.~Mirza}
\affiliation{Department of Physics, Isfahan University of
Technology, Isfahan 84156-83111, Iran}
\affiliation{Institute of Theoretical Science, University of Oregon, Eugene, OR 97403-5203 USA}
\author{M.~Zarei}
\affiliation{Department of Physics, Isfahan University of
Technology, Isfahan 84156-83111, Iran}

\date{\today}

\begin{abstract}
The effects of the IR aspects of gravity on quantum mechanics is
investigated. At large distances where due to gravity the
space-time is curved, there appears nonzero minimal uncertainty
$\Delta p_{0}$ in the momentum of a quantum mechanical particle.
We apply the minimal uncertainty momentum to some quantum
mechanical interferometry examples and show that the phase shift
depends on the area surrounded by the path of the test particle .
We also put some limits on the related parameters. This prediction
may be tested through future experiments. The assumption of minimal uncertainty in momentum can also explain the
anomalous excess of the mass of the Cooper pair in a rotating thin superconductor ring.
\end{abstract}



\maketitle

\newpage
\section{\large Introduction}
The quantum mechanics (QM) on a Riemannian curved space-time
background is the simplest part of the fundamental problem
associated with general relativity and the quantum world. On the other hand it is not possible to detect the classical gravitational effects on the quantum mechanical experiments unless during the interferometry experiments with neutrons as discussed in \cite{ab2}. In this paper we investigate another possibility to study the effects of gravity on quantum mechanical systems using minimal uncertainty in momentum assumption. It is
known that for large distances, where the curvature of space-time
becomes important, there is  no notion of a plane wave on a
general curved space-time \cite{kempf}. This means that there
appears a limit to the precision with which the corresponding
momentum can be described. One can express this as a nonzero
Minimal Uncertainty in Momentum (MUM) measurement. The minimal
uncertainty in momentum appears as an infrared (IR) effect of
gravity on a quantum mechanical system.
On the other hand, it has
been discussed in the literature that a minimal uncertainty in
position will be inevitable when a test particle tries to resolve
short distances \cite{kempf}.
 In order to probe the short
distances of the order of Planck length, $\ell_{P}$, test
particles require very high energies. According to Einstein's
equation, the gravity effects of high energy test particles must
significantly disturb the space-time structure when be probed.
Because of this phenomena, one expects a finite limit to the
possible resolution of distances. Therefore, due to ultraviolet
(UV) effects of gravity, quantum mechanics experiences a minimal
uncertainty in position. Also it has been shown that such a modified Heisenberg relation which leads to the minimal uncertainty in
position can be obtained from assuming a noncommutative structure for the geometry of space-time also leads to the
minimal uncertainty in
position \cite{g2}.
In one dimension, the minimal uncertainty
in position and momentum can be generalized in quantum mechanics
as follows \cite{kempf}
 \beq \Delta x\Delta p\geq \frac{\hbar}{2}(1+\alpha(\Delta x)^{2}+\beta(\Delta p)^{2}+\gamma)\label{modh} \eeq
where $\alpha$, $\beta$ and $\gamma$ are positive and independent
of $\Delta x$ and $\Delta p$. While in ordinary QM,the
uncertainty $\Delta p$ can be made arbitrarily small by letting
$\Delta x$ grow correspondingly, this is no longer the case in
the modified Heisenberg uncertainty relation (\ref{modh}). The
equation (\ref{modh}) can be deduced from the commutation
relation of the form \beq
[\hat{x}_{i},\hat{p}_{j}]=\frac{\hbar}{i}\:\delta_{ij}(\:1+\alpha
r^{2}+\beta p^{2})\label{com} \eeq with
$\gamma=\alpha<\textbf{x}>^{2}+\beta<\textbf{p}>^{2}$ and
$r^{2}=\sum x_{i}^{2}$. In order to have a closure form for the commutation relations, the equation (\ref{com}) is extended to the following form
\bea && [\hat{x}_{i},\hat{x}_{j}]\neq 0 \:\:\:\:\:\:  [\hat{p}_{i},\hat{p}_{j}]\neq 0 \eea
where the commutator $[\hat{x}_{i},\hat{x}_{j}]\neq 0$ implies a noncommutative structure for space-time.
The UV aspects of this UV/IR effects of gravity on quantum
mechanics, i.e., minimal uncertainty in position, have been
investigated widely in previous years \cite{minimal}. Also it has
been shown recently that this may be testable in a STM device
\cite{universal}.
At first, we concentrate on the IR aspects
of gravity and its effects on quantum mechanics. So, we assume
that in (\ref{modh}),  $\beta=0$  and study the quantum mechanics only with minimal
uncertainty in momentum. The Heisenberg uncertainty principle has
been verified experimentally for the fullerence molecules in
\cite{zeilinger}. We used the data generated in this work and fit
the generalized relation (\ref{modh}) for $\beta=0$ into these
data and found the following bound on

$\alpha$ \beq \sqrt{\alpha}\leq 3\times
10^{6}\:\texttt{m}^{-1}.\eeq

\noindent Also, it is possible to apply the MUM assumption to
other quantum mechanical examples for constraining the parameter
$\alpha$.
In order to construct an effective quantum theory and compare it with experimental data, similar to the minimal length theories \cite{ml}, we redefine the $\hat{x}$ and $\hat{p}$ in terms of the coordinate and momentum which satisfy the usual Heisenberg uncertainty relation. Then the whole effect of the MUM is transferred to a
new effective Schr\"{o}dinger equation which has terms dependent on the parameter $\alpha$. Therefor we find an effective quantum mechanical theory with corrections coming from the MUM assumption.
The generalized Heisenberg algebra with nonzero
$\alpha$ and $\beta$ can be represented on a position space wave function
$\psi(x)= <x|\psi>$ by letting $x$ and $p$ act as operators.

 \bea &&\hat{p}_{i}\cdot\psi(x)=
\frac{\hbar}{i}(1+\alpha r^{2})\partial_{i}\psi(x)\label{momentum}\\
&&\hat{x}_{i}\cdot\psi(x)=x_{i}\psi(x)\eea

\noindent It must be noticed that here the $x$'s are noncommutative coordinates.
Thus, in order to study each quantum mechanical problem
with the minimal uncertainty in momentum, it is sufficient to modify the Schr\"{o}dinger equation to a noncommutative one and
replace the momentum operator with the modified one given in
(\ref{momentum}).

 Here, we study the effects of minimal uncertainty momentum on
well understood quantum mechanic phenomena, such as, the
Aharonov-Bohm effect \cite{ab}, The Aharonov-Casher effect
\cite{ac}, COW effect \cite{cow}, flux quantization in
superconductors \cite{feyn}, rotating SQID \cite{squid}, Sagnac
effect \cite{page}, and gravitational AB effect \cite{gab} and the
Hydrogen atom. It will be shown that MUM makes a universal area
dependent correction in all these phenomena. This correction
disturbs the topological properties of these effects and may be
observed by suitably constructed experiments.

\section{\large The Aharonov-Bohm effect}

In 1959, Aharonov and Bohm  proposed an experiment to explore the
effects of the electromagnetic potential in the quantum domain
\cite{ab}. The standard configuration which was considered is the
interferometer pattern of the two slit diffraction experiments
involving a magnetic flux enclosed by two charged particle beams
to detect the phase shift. An excellent agreement was found
between the measured phase shift and the theoretical prediction

\beq \delta\varphi_{0}=\frac{q}{\hbar}\oint \textbf{A}\cdot d\ell
\eeq

\noindent where $q$ is the charge of particle, $\textbf{A}$ is the
vector potential, and the contour encloses the magnetic flux of a
solenoid. In this section, we calculate the modification in the
phase shift due to MUM by two quantum mechanics and semi-classical
approaches.

\subsection{ The quantum mechanics approach}

In noncommutative space the Schr\"{o}dinger equation can be written as \cite{ncsch}
\beq H\star |\psi>=E|\psi>\label{schh}
 \eeq
where the $\star$-product is defined as
\beq   (\textbf{A}\star\textbf{B})(x)=\textbf{A}(x_1)e^{\frac{i}{2}\theta^{ij}\partial^{(1)}_{i}
\partial^{(2)}_{j}}\textbf{B}(x_2)|_{x_1=x_2=x} \eeq

The Schr\"{o}dinger equation (\ref{schh}) in the presence of a vector potential is
\beq H\star\psi=\frac{1}{2m}\mathcal{D}_{j}\star \mathcal{D}_{j}\star\psi=k_{j}k_{j}\:\psi \label{eqqqq} \eeq where the variables $k_{j}$ are the eigenvalue of operators
$ \mathcal{D}_{j}=\hat{p}_{j}-qA_{j} $,
\beq \mathcal{D}_{j}\star \psi=k_{j}\psi \label{eqqq} \eeq
Here $A_j$ is the vector potential associated with magnetic field
$\textbf{B}$, $q$ is the charge of test particle and the operator $\hat{p}$ is the $\alpha$-dependent momentum operator defined in (\ref{momentum}).
One can solve this equation by choosing $\psi$ as
\beq \psi=e^{\Lambda}\eeq
Now we assume that noncommutativity is small then the equation (\ref{eqqq}) can be solved by perturbative expansion of $A_{j}$ and $\Lambda$ as follows
\bea && A_j=A^{(0)}_{j}+\theta A^{(1)}_{j}+\cdot\cdot\cdot \\&& \Lambda=\Lambda^{(0)}+\theta\Lambda^{(1)}+\cdot\cdot\cdot\eea
Then in the two slit experiments the wave function $\psi$ of charged particle
satisfies the following equation

 \bea
\mathcal{D}_{j}\star\psi &&\!\!\!\!\!\!\!\!=-i\hbar(1+\alpha r^{2})\partial_{j}e^{\Lambda}-qA_{j}\star e^{\Lambda}
\nonumber \\ && \!\!\!\!\!\!\!\!=e^{\Lambda}[-i\hbar(1+\alpha r^{2})\partial_{j}\Lambda -qA_j+\frac{iq}{2}\theta^{lm}\partial_{l}A_{j}\partial_{m}\Lambda\nonumber \\ &&\!\!+\mathcal{O}(\alpha\theta,\theta^{2})]=k_j\psi
 \label{sch}\eea
Then in order to the Schr\"{o}dinger equation be gauge invariant, $\Lambda$ must satisfy the following equation
\beq -i\hbar(1+\alpha r^{2})\partial_{j}\Lambda-qA_j+\frac{iq}{2}\theta^{lm}\partial_{l}A_{j}\partial_{m}\Lambda =k_j\eeq
Now for the first order of $A_j$ and $\Lambda$ one find
\beq -i\hbar(1+\alpha r^{2})\partial_{j}\Lambda^{(0)}-qA^{(0)}_j=k_j \label{eqq}\eeq
 which can be solved by a simple computation
\beq \partial_{j}\Lambda^{(0)}=\frac{iq}{\hbar}\frac{1}{1+\alpha
r^{2}} (A^{(0)}_{j}+k_j)\approx (\frac{iq}{\hbar }
-\frac{iq\alpha}{\hbar }r^{2})( k_j+A^{(0)}_{j}) \eeq Therefor
\bea \Lambda^{(0)}&&\!\!\!\!\!\!\!\!\!=\frac{iq}{\hbar }\int_{{\bf \ell}_{1}}^{{
\bf\ell}_{2}}{\bf k}\cdot d{\bf \ell}- \frac{iq\alpha}{\hbar
}\int_{{\bf \ell}_{1}}^{{\bf \ell}_{2}}r^{2}{\bf k}\cdot d{\bf
\ell}\:\nonumber\\&&
+\frac{iq}{\hbar }\int_{{\bf \ell}_{1}}^{{
\bf\ell}_{2}}{\bf A}\cdot d{\bf \ell}- \frac{iq\alpha}{\hbar
}\int_{{\bf \ell}_{1}}^{{\bf \ell}_{2}}r^{2}{\bf A}\cdot d{\bf
\ell}
\eea
where the terms containing $k_{j}$ is the free particle solution in the absence of magnetic field and will be discussed in the section 4. The terms containing $A_j$ gives the Aharonov-Bohm phase shift $\varphi_{AB}$.
For a closed path the phase $\varphi_{AB}$ is
corrected as \bea  \delta\varphi_{AB}
\!\!\!\!\!\!\!\!&&=\delta\varphi_{0}+\delta\varphi_{1}=\frac{q}{\hbar
}\int{\bf B}\cdot d{\bf s}-\frac{q\alpha}{\hbar }\oint r^{2}{\bf
A}\cdot d{\bf \ell}\nonumber \\&& =\frac{q}{\hbar }\Phi
-\frac{q\alpha}{\hbar }\int{\bf\nabla}\times (r^{2}{\bf A})\cdot
\hat{\bf n}ds \label{surf}
 \eea
which for a finite-radius solenoid, the vector potential for the
regions inside and outside of the solenoid are given by \bea {\bf
A}_{in}=\frac{B}{2}(-y,x,0) \label{in}\\{\bf
A}_{out}=\frac{B}{2}\frac{a^{2}}{x^{2}+y^{2}}(-y,x,0).\label{out}
\eea where, $a$ is the radius of the solenoid. Using this, the
surface integral in (\ref{surf}) for both inside and outside
regions of the solenoid is written as \bea &&\int{ \bf\nabla}\times
(r^{2}{\bf A})\cdot \hat{n}ds =2\int_{in}({\bf r}\times{\bf
A})\cdot\hat{n}ds\nonumber \\ &&+\int_{in}r^{2}{\bf\nabla}\times{\bf
A}\cdot\hat{n}ds+ 2\int_{out}(\vec{r}\times{\bf A})\cdot\hat{n}ds
\nonumber \\ &&+ \int_{out}r^{2}{\bf\nabla}\times{\bf A}\cdot\hat{\bf n}ds
\label{integral}.\eea Inserting (\ref{in}) and (\ref{out}) into
 (\ref{integral}), one finds \beq
\delta\varphi_{1}=-\frac{q}{\hbar
}\Phi\left(\alpha\frac{S_{\small out}}{\pi}-\alpha\frac{2}{\pi
a^{2}}\int_{in}r^{2}ds\right) \eeq where, provided that the radius
of solenoid, $a$, is small, the second term becomes negligible.
Consequently,  $S_{\small out}$ can be approximated by $S$, the
total area of the surface bounded by closed path. Therefore, the
$\varphi_{AB}$  can be written as \beq \delta\varphi_{\small
AB}=\frac{q}{\hbar}\Phi\left(1-\alpha\frac{S}{\pi}\right).
\label{faz}\eeq The second term in the phase shift causes the AB
effect to lose its topological properties. Before discussing the
phenomenological importance of (\ref{faz}), we will derive this
result using a semiclassical method in the next subsection. The parameter $\alpha$ appeared on the right hand of (\ref{faz}) may be dependent to the cosmological constant. The example that we have studied here is 2+1 dimensional systems living on
$\Re^{(2)} -\{0\}$-manifold. The use of three dimensional gravity has been suggested as a
test bed for the quantization of gravity \cite{witten}. Due to the smaller number of dimensions, this theory has tremendous
mathematical simplicity. The Einstein theory of gravity
in 2 + 1 space-time dimensions has a well know result,
namely, the  only nontrivial solution for the Einstein's equation is the
de-Sitter (or anti de Sitter) one and, therefore, the $\alpha$
coefficient should correspond to the cosmological
constant. Then the $\alpha$-dependent part of the equation (\ref{eqqqq})
actually, is the Schrodinger one in a de Sitter background. The AB effect on the anti de Sitter background has been studied in \cite{ads}.
If one retain the terms of the order of noncommutative parameter $\theta$, the following equation is found
\beq -i\hbar(1+\alpha r^{2})\partial_{j}\Lambda^{(1)}-qA^{(1)}_j+\frac{iq}{2}\theta^{lm}\partial_{l}A^{(0)}_{j}\partial_{m}\Lambda =0\eeq
which by integrating gives the $\Lambda^{(1)}$ term contributing the noncommutative effects in the AB phase shift. Here do not treat with this equation more and the solution can be found in \cite{ncsch} and \cite{chai}. In the following we ignore the $\theta$-dependent terms and only consider the $\alpha$-dependent corrections to the ordinary phase shifts.

 \subsection{ The semi-classical approach}

 In an interferometry experiment, the wave function of two spatially separated beams,
$\psi_{1}(\vec{r},t)=\phi_{1}({ r})e^{i\omega t}$ and
$\psi_{2}({\bf r},t)=\phi_{2}({\bf r})e^{i\omega t}$ can be
described in terms of particle trajectories. In this
semi-classical approximation, $\phi_{i}({\bf r})$ is written in
the following form \cite{ab2},
 \beq \phi_{i}({\bf r})=\sqrt{\rho_{i}({\bf r})}\:\exp\left(\frac{i\mathcal{S}_{i}({\bf r})}{\hbar}\right)\label{eq}\eeq
Where, $\mathcal{S}$ can be identified with the classical action.
Substitution of (\ref{eq}) in the minimal momentum Schr\"{o}dinger
equation will show that $\mathcal{S}$ obeys the following minimal
momentum eikonal equation \beq \left((1+\alpha
r^{2})\nabla\mathcal{S}_{i}\right)^{2}=\textbf{p}^{2}_{c}\label{eikonal}
\eeq where, $\textbf{p}_{c}$ is the canonical momentum. Then,
equation (\ref{eikonal}) can easily be solved and the wave
function will be as \bea \phi_{i}({\bf
r})\!\!\!\!\!\!\!\!&&=\sqrt{\rho_{i}({\bf
r})}\:\exp\left(\frac{i\mathcal{S}_{i}({\bf
r})}{\hbar}\right)\nonumber \\&& \equiv \phi_0
\exp\left(\frac{i}{\hbar}\oint\frac{ {\bf p}_{c}\cdot d{\bf
\ell}}{1+\alpha r^{2}}\right) \eea where, $\phi_{0}$ is the
unperturbed part of the wave function. In the AB experiment, the
integration encloses the interferometry surface and ${\bf p}_{c}$
is defined by the Lagrangian of a charged particle in a magnetic
field \beq \mathcal{L}=\frac{p^{ 2}}{2m}+q{\bf v}\cdot {\bf A}
\eeq From this, the canonical momentum will be given by
$\textbf{p}_{c}=m\dot{\textbf{r}}+q\textbf{A}$. The phase of the
perturbed part of the wave function will then be given by
\bea\mathcal{S}\!\!\!\!\!\!\!\!\!&&=q\oint\frac{{\bf A}\cdot
d\ell}{1+\alpha r^{2}}\nonumber \\&&\approx q [ \oint{\bf A}\cdot
d\ell-\alpha\oint r^{2}{\bf A}\cdot d\ell ] \eea Therefore, the
correction to the Aharonov-Bohm phase is calculated as \beq
\delta\varphi_{\small
AB}=\mathcal{S}/\hbar=\delta\varphi_{0}\left(1-\alpha\frac{S}{\pi}\right).
\eeq where $\delta\varphi_{0}=\frac{q}{\hbar }\Phi$. An
estimation for the upper bound on the parameter of $\alpha$ can
be made using the available experimental data on the
Aharonov-Bohm effect. The contribution of the extra phase shift
due to MUM related to the usual shift of phase is \beq
\left|\frac{\delta\varphi_{1}}{\delta\varphi_{0}}\right|=\frac{\alpha
S}{\pi}\label{ratio} \eeq Fitting the ratio (\ref{ratio}) into
the accuracy bound of the experiments to verify the AB effect,
one obtains a bound on the parameter $\alpha$. The experiment
reported in \cite{abex1}  with an error of $11\%$ gives a
constraint on $\alpha$ as follows \beq \sqrt{\alpha}\leq 6\times
10^{2}\: \texttt{m}^{-1} \eeq where, we have estimated the area
surrounded by two electron beams as $S\approx
1\:\mu\texttt{m}^{2}$.

\section{\large The Aharonov-Casher effect}

Following the semi-classical approach of the previous section, we
study the Aharonov-Casher (AC) effect \cite{ac} by assuming a
minimal uncertainty in momentum and find variation in the AC
phase shift $\delta\varphi_{AC}$. The AC effect is the dual of the
AB effect. The AC phase will  accumulate when a test particle
carrying a magnetic moment ${\bf\mu}$ travels around a charged
wire. It is simple to verify that the canonical momentum for the
MUM condition is given by \beq {\bf
p_{c}}=m\dot{\textbf{r}}+\frac{1}{1+\alpha r^{2}}\:{\bf
\mu}\times{\bf E}
   \eeq  where, the electric field ${\bf E}$ is
\beq {\bf E}=\frac{\lambda}{2\pi r}\:\hat{r} \eeq where, $\lambda$
is the charge per unit length. The phase shift for a test particle
diffracting around the line charge is calculated by \bea
\delta\varphi_{AC}\!\!\!\!\!\!\!\!&&=
\frac{1}{\hbar}\oint\frac{{\bf p_{c}}\cdot d{\bf \ell}}{1+\alpha
r^{2}}=\frac{\lambda}{2\pi\varepsilon_{o}\hbar}\oint\frac{{\bf
\mu}\times{\bf E}\cdot d{\bf \ell}}{r^{2}(1+\alpha
r^{2})}\nonumber\\ && \approx
\frac{\lambda}{2\pi\varepsilon_{o}\hbar}\oint\frac{{\bf
\mu}\times{\bf r}\cdot d{\bf
\ell}}{r^{2}}-\alpha\frac{\lambda}{2\pi\varepsilon_{o}\hbar}\oint{\bf
\mu}\times{\bf r}\cdot d{\bf \ell}\nonumber\\
&&=\delta\varphi_{0}-\alpha\frac{\lambda}{2\pi\varepsilon_{o}\hbar}\int\nabla\times({\bf
\mu}\times{\bf r})\cdot\hat{\bf n}\:ds \eea Therefore, \beq
\delta\varphi_{AC}=\delta\varphi_{0}\left(1-\frac{\alpha
S}{\pi}\right)\label{acphi} \eeq where,
$\delta\varphi_{0}=\lambda\mu$. Similar to the previous section
the experimental observations on the AC phase shift can be used
to put a limit on the $\alpha$ parameter. In the experiment
described in \cite{acexp}, the area can be approximated as
$S\approx 4 \:\texttt{cm}^{2}$. Then, fitting (\ref{acphi}) into
the accuracy bound of this experiment which is $24\%$, one obtain
\beq \sqrt{\alpha}\leq 0.5\times 10^{2}\:\texttt{m}^{-1}.\eeq
which is near to the AB case. The effect of noncommutativity on the AC phase shift has been studied in \cite{ncsch}

\section{\large The COW effect }
The MUM can affect the results of a experiment was concerned by
Collela, Overhauser and Werner (COW) \cite{cow} in 1975 with
neutron interferometry in gravitational field of the Earth. In
this experiment a beam of neutron is split  into two parts, such
that they can travel at different heights in the gravitational
field of the Earth with different velocities. Using the first
term of canonical momentum, $m\dot{\textbf{r}}$, the phase shift
in the interferometer experiment in a situation where the neutron
go through the loop ABCD, is given as \beq
\delta\varphi_{0}=\frac{m}{\hbar}\oint \textbf{v}\cdot
d\ell\approx \frac{m(v_{0}-v_{1})}{\hbar}\overline{AB} \eeq where
$v_{0}$ and $v_{1}$ denote the velocities along the paths
$\overline{AB}$ and $\overline{CD}$. Then, following the
discussion of previous sections, the MUM modification phase shift
is given by \beq
\delta\varphi=\delta\varphi_{0}\left(1-\alpha\frac{S}{\pi}\right)
\eeq and using the 1\% accuracy confirmed by COW experiment, the
following bound on $\alpha$ is obtained \beq  \sqrt{\alpha}\leq
0.5\times 10 \:\texttt{m}^{-1} \eeq


\section{\large The flux quantization }
It is well known that the magnetic flux passing through any area
bounded by a superconducting ring is quantized \cite{feyn}. The
quantization of magnetic flux is closely related to the
Aharonov-Bohm effect. The quantum of this magnetic flux is
universal, independent of the ring properties and equals to \beq
\Phi_{0}=\frac{\pi\hbar}{e}\approx 2\times
10^{-7}\:\texttt{gauss-cm} \eeq But this value changes when we
consider the MUM assumption. Under such conditions, the flux
quantization is modified as \beq
\Phi=\frac{q}{\hbar}\oint\frac{\textbf{ A}\cdot d\ell}{1+\alpha
r^{2}} \eeq The only physical requirement is that there can be
only one value of the wave function \beq
\psi=\sqrt{\rho}e^{i\frac{q}{\hbar}\oint\frac{\textbf{ A}\cdot
d\ell}{1+\alpha r^{2}}} \eeq on a closed path. Then the flux
becomes \beq \Phi\approx\frac{\pi n\hbar}{e}\left(1-\frac{\alpha
S}{\pi}\right) \eeq where, $S$ is the horizontal area bounded by
supercurrent (superconducting electrical current). Hence, the
quantum of the magnetic flux is changed as \beq
\Phi_{0}\approx\frac{\pi \hbar}{e}\left(1-\frac{\alpha
S}{\pi}\right) \eeq Among the experiments for measuring $\Phi_0$
which can be used to constrain $\alpha$,  paper \cite{fexp} has
found that the quantum of the magnetic flux trapped in a hollow
superconductor is  $\pi\hbar/e\pm 4\%$. In this experiment, the
area is $S\approx 3\times 10^{-8}\:\texttt{cm}^{2}$,  therefore,
we will have \beq \sqrt{\alpha}\leq 4\times
10^{5}\:\texttt{m}^{-1}\eeq which is near to the bound obtained in
Section 1 by fitting with data from the experiment verifying the
Heisenberg uncertainty relation.

\section{\large The Sagnac effect }

The Sagnac effect \cite{page} for light waves is also valid for
matter waves and it has been verified experimentally \cite{ab2}.
In the well known Sagnac effect, an extra shift
$\delta\varphi_{S}$ arises when observing the interference
between two beams in the flat space-time along a closed path due
to the rotation of the interferometer. In this section, we study
the effect of the rotating frame on an electron wave
interferometry experiment. Similar to the previous section, the
eikonal approximation is used in combination with the assumption
that the interferometer is small in comparison to the radius of
the Earth. Using the Lagrangian of a charged particle on rotating
Earth, one can find the canonical momentum ${\bf p_{c}}$ as \beq
{\bf p_{c}}=m\dot{{\bf r}}+m\bf{\omega}\times{\bf r} \eeq where,
$m$ is the mass of particle and ${\bf \omega}$ is the angular
velocity. The phase shift induced by the rotation of frame is
given by \beq \delta\varphi_{S}=\frac{m}{\hbar}\oint {\bf
\omega}\times{\bf r}\cdot d\ell \eeq
 If one assumes the MUM condition to hold, then phase shift is modified to the following form
\beq \delta\varphi_{S}=\frac{m}{\hbar}\oint \frac{{\bf
\omega}\times{\bf r}\cdot d\ell}{1+\alpha
r^{2}}=\delta\varphi_{0}-  \frac{m\alpha}{\hbar}\oint r^{2}{\bf
\omega}\times{\bf r}\cdot d\ell\eeq The calculation of the second
term gives the following solution \beq
\delta\varphi_{1}=-\frac{m\alpha}{\hbar}\oint r^{2}{\bf
\omega}\times{\bf r}\cdot d\ell=-\frac{m\gamma}{\hbar}\int
\nabla\times[r^{2}{\bf \omega}\times{\bf r}]\cdot{\bf \hat{n}} ds
\eeq For horizontally incident beams,  $\delta\varphi^{(1)}$
becomes \beq \delta\varphi_{1}=-\frac{4 m}{\hbar}\:\alpha\int
r^{2}ds\approx
-\delta\varphi_{0}\alpha\left(\frac{2S}{\pi}\right) \eeq where,
$\delta\varphi_{0}=(2m/\hbar)\omega S$. The dependence on surface
indicates that the Sagnac effect is not as topological as the
Aharonov-Bohm effect. The total Sagnac phase shift is given by
\beq
\delta\varphi_{S}=\delta\varphi_{0}\left(1-\alpha\frac{2S}{\pi}\right)
\eeq In reference \cite{sagexp}, the phase shift caused by
rotation of an electron biprism interferometer placed on a
turntable has been measured for different areas. In the
experimental setup with $\omega/2\pi=0.5\:\texttt{s}^{-1}$, the
area $S\approx 2.8\texttt{ mm}^{2}$ and the error of about
$30\%$, the parameter $\alpha$ is bounded as \beq
\sqrt{\alpha}\leq 4\times 10^{2}\:\texttt{m}^{-1}. \eeq For
different areas, this bound will have the same order of magnitude.

\section{\large Rotating superconductor }
An interesting investigation of the Sagnac effect can be made in a
rotating superconducting system by use of  Superconducting
Quantum Interference Devices (SQUID) \cite{squid}. We consider a
SQUID involving interference between current flow through a pair
of Josephson junctions rotating with a constant angular speed
$\omega$ about the $z$ axis perpendicular to the SQUID. Now if
$x_{\mu}$ denotes the coordinates of the initial framework and
$x'_{\mu}$ denotes the coordinates of rotating framework, then
one can verify that \bea x=x'\cos\omega t'-y'\sin\omega t' \\
y=x'\sin\omega t'+y'\cos\omega t'\eea and the Lagrangian for
motion in the rotating frame becomes \beq
\mathcal{L}=-m(g_{\mu\nu}\dot{x}'^{\mu}\dot{x}'^{\nu})^{1/2} \eeq
where, $g_{\mu\nu}$ is defined as \beq
g_{\mu\nu}(x')=\frac{\partial x^{\alpha}}{\partial
x'^{\mu}}\frac{\partial x^{\alpha}}{\partial
x'^{\nu}}\eta_{\alpha\beta} \eeq where, $\eta_{\alpha\beta}$ is
the Minkowski metric. The $g_{\mu\nu}$ component is given by \bea
&&
g_{00}=1-(\omega\times\textbf{x}')\cdot(\omega\times\textbf{x}')
\nonumber \\ && g_{0i}=(\textbf{x}'\times\omega)_{i},\:\:\:
g_{ij}=\delta_{ij} \eea and therefore $
\mathcal{L}=-m+\frac{1}{2}m\dot{{\bf
x}}'^{2}+e\tilde{\phi}-e\dot{{\bf x}}'\cdot\tilde{{\bf A}} $,
 where the effective scalar and vector potentials $\tilde{\phi}$
and $\tilde{{\bf A}}$ due to rotation are defined as \bea &&
\tilde{\phi}=\frac{m}{2e}(\omega\times\textbf{x}')\cdot(\omega\times\textbf{x}')\nonumber
\\ &&\tilde{{\bf A}}=-\frac{m}{e}(\textbf{x}'\times\omega)\eea
and the angular velocity $\omega$ can be interpreted as a
magnetic $B$ field which becomes zero within the rotating
superconductor. This leads to a phase shift $\delta\varphi_{0}$
in the interference pattern between two supercurrents flowing
through different paths in the SQUID. For the MUM condition, the
phase shift is calculated as follows \beq
\delta\varphi=\frac{2e}{\hbar}\oint\frac{\tilde{{\bf A}}\cdot
d\ell }{1+\alpha
r^{2}}\approx\delta\varphi_{0}\left(1-\alpha\frac{S}{\pi}\right)\eeq
where, $\delta\varphi_{0}$ is given by \beq
\delta\varphi_{0}=\frac{4m\omega S}{\hbar}\label{rot} \eeq and
$S$ is the area of nonsuperconducting region enclosed by two
supercurrents. An experiment on rotating superconductor has been
reported in \cite{squidexp} in which the phase shift (\ref{rot})
is  measured indirectly with an error of about $30\%$. In this
experiment, $S\approx 0.074\:\texttt{cm}^{2}$ and
$\omega=10\:\texttt{rad}\texttt{s}^{-1}$. With this information,
the bound on $\alpha$ is given as \beq \sqrt{\alpha}\leq 3.6\times
10^{2}\:\texttt{m}^{-1}.\eeq
It can also be possible to use the MUM assumption in order to interpret the Cabrera and Tate experiment \cite{tate} which reported an anomalous excess of mass of the Cooper pair in rotating thin superconductor ring. In this experiment the following deference between experimental Cooper mass, $m^{\ast}$, and it's theoretical prediction, $m$, reported as
\bea \Delta m&&\!\!\!\!\!\!=m^{\ast}-m =1.023426(21)\:\texttt{MeV}-1.002331\:\texttt{MeV}\nonumber\\&&\!\!\!\!\!\!=94.147240(21)\:\texttt{eV}\label{dm} \eea
So far this experimental result never received an explanation in the context of superconductor's physics. Hence it has been done some works in the context of quantum gravity or dark energy to interpret it \cite{matos}. For the case that MUM is assumed, one find that
\beq \frac{\hbar}{m^{\ast}}=2S\Delta\nu\left(1-\alpha^{\ast}\frac{S}{\pi}\right) \eeq
where $S$ is the area bounded by the closed path, $\Delta\nu$ is the flux null spacing and $\alpha^{\ast}$ is the constant of MUM defined in the rotating frame.
 As shown in \cite{ord}, the coefficient $\alpha$ can be dependent on the metric of space-time. Therefor one expect that $\alpha$ in a rotating and unrotating frame be different. Hence one can write
\beq \frac{\Delta m}{m}\approx S\Delta\alpha \eeq
where $\Delta m$ was defined in (\ref{dm}) and $\Delta\alpha=\alpha^{\ast}-\alpha$. So MUM can give an alternative interpretation for the anomalous Cooper pair excess. From the numerical values of the Cabrera and Tate experiment, we find $\Delta\alpha$ as
\beq \Delta\alpha\approx 5\times 10^{-2}\:\texttt{m}^{-2}.\eeq

\section{\large The gravitational Aharonov-Bohm effect}

In this section, we will consider the minimal uncertainty in
momentum in the gravitational analog of Aharonov-Bohm effect. This
teleparallel approach to gravitation consists in a  shift similar
to the Aharonov-Bohm effect, but produced by the presence of a
gravitational gauge potential. This analog phase shift is zero in
 the Newtonian gravity. In general relativity, both mass and angular
momentum act as the source of gravity but in the Newtonian gravity
only the mass, not the angular momentum, gravitates. These two
degrees of freedom are analog to the charge and magnetic moments
in an electromagnetic field.
Therefore, the analog of the AB effect for gravity is the phase
shift $\delta\varphi$ acquired by a mass going around a string or
rod that has angular momentum.
 We now consider the linearized limit of general relativity and
 write the metric as $g_{\mu\nu}=\eta_{\mu\nu}+h_{\mu\nu}$
 where, $h_{\mu\nu}\ll 1$
and  $\vec{h}_0=-(h_{01},h_{02},h_{03})$ are the Coriolis
   vector potentials.
   The phase shift due to
   gravitational fields in a situation where a particle with mass
   $m$ and spin $S$ interferes around a cylinder is given by
 \beq \delta\varphi=-\frac{1}{2\hbar}\oint h_{\mu a,b}S^{ab}dx^{\mu} \eeq
 where, $S^{ab}$ $(a,b=0,1,2,3)$ is a generator for Lorentz transformation in the spin
  space related to the spin vector $S^{a}$ and the 4-velocity $v^{b}$ as
 \beq S^{ab}=\epsilon^{abcd}v_{c}S_{d}. \eeq
The above phase shift is the gravitational analog of AC phase
shift \cite{gab3} and can be simplified further to
 \beq \delta\varphi=-\frac{2}{\hbar}\oint{\bf g}\times{\bf S}\cdot d\ell\eeq where, $S$
 is the spin of the test particle and ${\bf g}$ is the
  acceleration due to gravity defined as ${\bf g}=-\frac{1}{2}\nabla h_{00}$.
The dual of this situation is the gravitational analog of the AB
effect \cite{gab2} which is given by \beq\delta\varphi=
\frac{m}{\hbar}\oint h_{0,\mu }dx^{\mu}\eeq This phase shift is
given through interfering a mass $m$ around the cylinder with an
angular momentum $\textbf{J}$  given by \beq
\vec{h}_0=-\frac{4G}{\rho_{0}^{2}}\textbf{J}\times\textbf{r} \eeq
where, $\rho_{0}$ is the radius of the cylinder and $G$ is the
Newton constant. The two phase shifts obtained above can also be
derived through a Lagrangian of the form, \beq
\mathcal{L}=\frac{1}{2m}(\textbf{p}-m\vec{h}_0-2\textbf{S}\times\textbf{g})^{2}+\frac{1}{2}mh_{00}
\eeq
 Using this Lagrangian, one can extend the phase shifts for the case
  of minimal uncertainty in momentum condition. In a semi-classical situation, the modified phase
  shifts
  are given by
\bea
\delta\varphi_m\!\!\!\!\!\!\!\!&&=\delta\varphi^{(1)}+\delta\varphi^{(2)}
\nonumber \\&&=-\frac{2}{\hbar}\oint\frac{{\bf g}\times{\bf
S}\cdot d\ell}{1+\alpha r^{2}}+\frac{m}{\hbar}\oint
\frac{\vec{h}_{0}\cdot d\ell}{1+\alpha r^{2}} \eea  whose
calculations are straightforward and lead to  results similar to
those in  the previous sections.
The AB phase shift is given by a solenoid producing a vector potential which satisfy the Maxwell's equation. Corresponding to this solenoid, for the Einstein's equation there is a spinning cosmic string solution which has angular momentum and mass. A gravitational AB phase shift can be produced by such cosmic string i.e. the solenoid may be replaced by a cosmic string. Outside the cosmic string, the curvature and torsion vanish,
which is analogous to the vanishing of the electromagnetic field strength outside
the solenoid. Then a quantum mechanical particle with the state $|\psi>$ enclosing the cosmic string.
Another interesting issue is to consider the AB scattering of cosmic string with a quantum charge particle. The AB effect as a mechanism for
detecting cosmic strings \cite{cosmicS}. This effect is similar to the case of a charge particle scattering off an infinitesimally thin solenoid. The interaction of cosmic string with a relativistic quantum particle, the differential cross section is given by writing the Dirac equation and then calculating the scattering amplitude. The MUM assumption will modify the AB cross section into a cross section with $\alpha$-corrections.

\section{\large The Hydrogen atom}
The effects of MUM can also be studied at the atomic scales. We
have calculated the effects of MUM on Hydrogen atom where energy
levels are modified as \beq E_{n,l}=
E^{(0)}_{n,l}+\frac{(\alpha\hbar)^{2}}{2m_{e}}\left((n+l)^{2}+l(l+1)-\frac{l(l+1)}{n}\right)\eeq
where, $E^{(0)}_{n,l}$ is the energy related to the usual Hydrogen
system. In order to obtain a bound on $\alpha$, we use the
measurements on the Hydrogen energy levels similar to what has
been done in paper \cite{hyd}.  In that paper, the Hydrogen
$\texttt{1S-2S}$ transition frequency, $\omega$, has been measured
with an accuracy of $1.8$ parts in $10^{14}$. Using this
measurement, one has \beq \frac{\delta \omega}{\omega}\leq
1.8\times10^{-14}\eeq Then, the parameter $\alpha$ is constrained
as \beq \sqrt{\alpha}\leq 1.6\times 10^{2}\:\texttt{m}^{-1}.\eeq

\section{\large Conclusion}

We argued that QM on a curved space-time experiences a minimal
uncertainty in momentum. This feature can be considered as the
effect of IR aspects of gravity on quantum mechanics. Based on
this assumption, we resolved some quantum mechanics examples such
as Aharonov-Bohm effect, Aharonov-Casher effect, COW effect, flux
quantization, Sagnac effect, rotating superconductor, and
gravitational Aharonov-Bohm effect and Hydrogen atom energy
levels. We found that the phase shift due to this effects is
corrected by an area dependent term. By fitting this new phase
shift with the experiments performed to verify the above effects
and also the data of experiment performed for exploring the
Heisenberg uncertainty relation, we found that the parameter
$\alpha$ is bounded as $\sqrt{\alpha}\leq
10-10^{6}\:\texttt{m}^{-1}$. In order to explore the area
dependent phase shift predicted in this paper, one can, for
instance prepare and do the interferometry experiment with
different area conditions similar to the setup used in
\cite{sagexp}. It was also shown that the assumption of minimal uncertainty in momentum can explain the
anomalous excess of the mass of the Cooper pair in a rotating thin superconductor ring in the Cabrera and Tate experiment.

\section*{\small Acknowledgment}

We would like to thank M. Haghighat and F. Loran for useful discussions.


\end{document}